\documentclass[prl,amssymb,twocolumn,showpacs,superscriptaddress]{revtex4}

\usepackage{graphicx}
\usepackage{bm}

\begin{document}

\title{Out-of-equilibrium dynamics of the vortex glass}
\author{Sebastian Bustingorry}
\affiliation{Centro At{\'{o}}mico Bariloche, 
8400 San Carlos de Bariloche,
R\'{\i}o Negro, Argentina.}
\author{Leticia F. Cugliandolo}
\affiliation{LPTHE, 4 Place Jussieu,  75252 Paris Cedex 05, France\\
and LPT Ecole Normale Sup{\'e}rieure, 24 rue Lhomond, 75231
Paris Cedex 05 France.}
\author{Daniel Dom\'{\i}nguez}
\affiliation{Centro At{\'{o}}mico Bariloche, 
8400 San Carlos de Bariloche,
R\'{\i}o Negro, Argentina.}

\begin{abstract}

We study the relaxational dynamics of
flux lines in high-temperature superconductors with random
pinning using Langevin dynamics.
At high temperatures the dynamics is stationary and the fluctuation
dissipation theorem (FDT) holds. 
At low temperatures the system does not equilibrate with 
its thermal bath: a simple multiplicative aging is
found, the FDT is violated and we found that an effective
temperature characterizes the slow modes of the system. 
The generic features of the evolution -- scaling laws -- are dictated
by the ones of the single elastic line in a random environment.

\end{abstract}


\maketitle

The effect of quenched disorder on
the vortex phase diagram of  superconductors
has attracted much interest 
since the discovery of high $T_c$ cuprates.
Since the long-range order of the vortex lattice is 
destroyed by any small amount of disorder \cite{larkin},
an amorphous ``vortex glass'' (VG) phase was predicted \cite{fisher}.
While it is currently understood that there is a ``Bragg glass'' phase
for weak disorder \cite{gld},
there is still
some controversy about the 
nature of the 
amorphous phase at
strong disorder. 
Experimentally, one finds an irreversibility line (IRL)
below which the magnetization
is irreversible and the linear resistivity drops to 
nearly zero \cite{irl}.
The original proposal that the IRL signals
a thermodynamic continuous transition from the 
vortex liquid (VL) to a VG  with zero linear resistivity \cite{fisher}
was initially supported  by experiments in which
a scaling of current-voltage (IV) curves 
was found \cite{koch} and by
simulations in randomly frustrated 3D XY models 
with a finite temperature transition \cite{gglass}.
However, when magnetic screening is included in these 3D XY models
the transition disappears \cite{bokil}.
Moreover, recent  experiments showed that the IV curves
do not scale  as expected \cite{strachan}. 

One possible scenario for the IRL is that at low temperatures
experiments occur out of equilibrium. 
Simulations of the London-Langevin model for vortices (which
includes screening) showed that 
below a crossover temperature, where the VG-type of 
criticality is arrested, the time scales grow very quickly, similar
to a Volger-Fulcher form \cite{zimanyi} as in structural glasses
\cite{angell}. 
Therefore, it
is important to compare the dynamic behavior of the vortex system
with that of structural glasses.
In this paper we apply the analysis of correlation and response functions
used in the study of relaxation in structural glasses \cite{lesuch} 
to analyze the VG phase in a simulation of a London-Langevin model
similar to the one studied in Ref.~\cite{zimanyi,zimanyi2}.
 
We consider 3D elastic flux lines in a superconductor where the $i$-th line
has coordinates ${\bf r}_i(z)=[x_i(z),y_i(z)]$ 
with $z$ the vertical direction along the  magnetic field $B$. 
The superconductor has anisotropy $\epsilon=\xi_c/\xi_{ab}=
\lambda_{ab}/\lambda_c$, with $\xi_{ab}$, $\xi_c$ the
coherence lengths and $\lambda_{ab}$, $\lambda_c$
the penetration depths; the axis are such that
$c\parallel z$ and $ab\parallel {\bf r}_i$.
Assuming that ${\bf r}_i(z)$ varies slowly with $z$,
the Hamiltonian for a London model of elastic flux lines is
$$
{\cal H}= \sum_{z} 
\left[\sum_i
U_{l}(\Delta{\bf r}_{iz})
+ U_{d}({\bf r}_{iz})
+ \sum_{i<j}U_{in}({\bf r}_{jz}-{\bf r}_{iz})
\right]
.$$
Having discretized $z$ in units of $d_z$, 
${\bf r}_{iz}\equiv{\bf r}_i(z)$
represents the 2D position of a line element in the
plane $z$.
The interaction energy between line elements in the
plane $z$ is approximated as 
$U_{in}= 2 \epsilon_0 d_z K_0(r/\lambda_{ab})$
with $\epsilon_0=(\Phi_0/4\pi\lambda_{ab})^2$ \cite{zimanyi,zimanyi2,ryu}.
The elastic line energy of the $i$-th
vortex  is $U_{l}= \frac{1}{2}c_{l} (\partial_z {\bf r}_i)^2
dz$, which is discretized as $U_{l}
= \frac{1}{2} c_{l} (\frac{\Delta {\bf r}}{d_z})^2 d_z$, with
$\Delta{\bf r}_{iz}={\bf r}_{i,z+1}-{\bf r}_{i,z}$.
A natural choice for  $d_z$ is the distance between CuO planes,
in which case one can account for the effect 
of the Josephson coupling \cite{ryu} using
$U_{l}=2c_{l}\lambda_J \frac{|\Delta {\bf r}|}{d_z}$ 
for  $|\Delta{\bf r}| > 2 \lambda_J$ (keeping the previous expression
for $|\Delta{\bf r}| < 2 \lambda_J$), 
where $\lambda_J = d_z/\epsilon$ is the 
Josephson length. 
The quenched disorder potential due to impurities
is $U_{d}({\bf r})=\int d^2{\bf r'} u({\bf r}')
p(|{\bf r}-{\bf r}'|)$, where $p(r)= 2\xi_{ab}^2/(r^2+2 \xi_{ab}^2)$
and $\langle u({\bf r},z) u({\bf r}',z') \rangle = \gamma\delta({\bf
r}-{\bf r}')\delta_{zz'}$ defines the disorder
strength $\gamma$   \cite{zimanyi2,blat94}.
We model the dynamics with the Langevin equation
$$\eta\frac{\partial {\bf r}_{iz}(t)}{\partial t} = -\frac{\delta {\cal
H}[\{ {\bf r}_{lz}(t)  \}]}{\delta {\bf r}_{iz}}+{\bf f}^T_{iz}(t) \; ,$$
where  $\eta$ is the Bardeen-Stephen friction coefficient. 
The thermal force ${\bf f}^T_{iz}(t)$ 
satisfies $\langle
f^T_{iz,\mu}(t) \rangle = 0,$ and $\langle f^T_{iz,\mu}(t)
f^T_{i'z',\mu}(t') \rangle = 2\eta k_BT \delta(t-t') \delta_{zz'}
\delta_{ii'}\delta_{\mu \mu'}$, where $\mu,\mu'=x,y$
and $T$ is the thermal bath temperature. 

The above model gives a good quantitative description
of the vortex physics of 
moderately anisotropic high-$T_c$ superconductors
like YBa$_2$Cu$_3$O$_{7-\delta}$ \cite{zimanyi2,ryu,blat94,bscco}.
We therefore choose parameter
values corresponding to YBa$_2$Cu$_3$O$_{7-\delta}$:  $\epsilon=1/5$, 
$\lambda_{ab}/\xi_{ab}=100$, and
$\lambda_J/\xi_{ab}=16$; and we use 
$c_{l}=\epsilon^2\epsilon_02\left[1+\ln
(\lambda_{ab}/d_z)\right]/\pi$ \cite{ryu}.
The strength of
disorder is set to $\gamma = 10^{-5}$, for which
case we find  
that above $B_{cr}\sim 0.002 H_{c2}$
the Bragg peaks dissapear and the flux lines
are  frozen in a highly amorphous structure
at low temperatures.
We therefore choose to study  the case with
$B=0.01 H_{c2} \gg B_{cr}$, which
is deep within the VG regime at low $T$.
Time is normalized by $t_0=\xi_{ab}^2
\eta/\epsilon_0$, length by the vortex lattice parameter
$a_0=[2\Phi_0/(\sqrt{3}B)]^{1/2}$, energy by 
$\epsilon_0d_z$, and temperature by $\epsilon_0d_z/k_B$. 
We simulate 
 $N=56$ vortices in a box of size
$7a_0\times8a_0\sqrt{3}/2$ 
with periodic boundary conditions for the
 in-plane coordinates. 
The $z$ direction is discretized in $L=50$
planes with free boundary conditions.  
Averages are performed over $10$ realizations
of the disorder.

In order to study the out-of-equilibrium dynamics we use the following
protocol. First we equilibrate the system at an initial 
high temperature well inside the VL 
($T_i=0.3$) evolving during $t=10^4$ steps.
Then we quench the system to a low temperature $T$,
where the time count is set to zero. Starting with this 
far from equilibrium initial condition, the 
system is let evolve during a
waiting time $t_w$, 
after which the quantities of interest are measured.
In general, 
one defines two-time correlation or response functions $C(t,t_w)$.
When the system reaches equilibrium, these quantities become independent
of $t_w$ and depend only on the difference $\tau=t-t_w$.
If the system is not able to reach equilibrium within the observation
time window, $C(t,t_w$) depends on the two times. In particular, if
the decay gets slower for longer $t_w$'s we say that the system
``ages''.
We study: (a) The dynamic
wandering \cite{yoshino}, $W(z,t,t_w)$, defined by 
\begin{figure}[!tbp]
\includegraphics[angle=-90,width=\linewidth,clip=true]{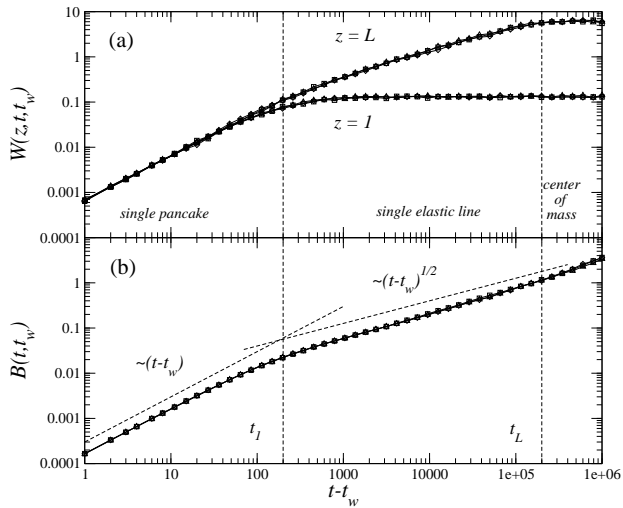}
\caption{\label{f:wyb}(a) Dynamic wandering and (b) mean square
displacement for elastic lines with no in-plane
interaction and without disorder, after the quench to
$T=0.06$. The elastic correlation times $t_1$ and
$t_L$ separate three regimes (single pancake, elastic
line and center of mass diffusion). The behaviors $B\sim(t-t_w)$
and $B\sim(t-t_w)^{1/2}$ are highlighted.}
\end{figure}
$$NW(z,t,t_w)=
\sum_i \langle |[\mathbf{r}_{iz}(t)-\mathbf{r}_{i0}(t)]-
[\mathbf{r}_{iz}(t_w)-\mathbf{r}_{i0}(t_w)]|^2\rangle$$
which measures how the displacement of the vortex segment in the $z$-th
plane with respect to the bottom plane 
correlates between $t$ and $t_w$ 
($\langle\ldots\rangle$ means average over thermal noise and disorder).
(b) The mean square displacement (MSD) in the planes, 
$$B(t,t_w)=\frac{1}{LN}\sum_{iz} \langle [x_{iz}(t)-x_{iz}(t_w)]^2\rangle.$$

Firstly, we analyze the single, non-interacting, flux line
without disorder. In Fig.~\ref{f:wyb}(a) and (b) we
show the dynamic wandering and the MSD, respectively, when
the line is quenched to $T=0.06$. In this case the flux line
reaches equilibrium in a short time:
both correlation functions are
independent of $t_w$  and depend only on $\tau=t-t_w$ 
(data for several $t_w$ overlap in the plots).
From the behavior of $W$,
we define a set of 
characteristic times $t_z$, such that
when $\tau=t-t_w > t_z$,
$W(z,\tau\gg t_z)\sim z^{2\zeta}$ saturates to a constant value.
Here $\zeta$ is the roughness exponent
given by thermal fluctuations, 
{\it i.e.} $\zeta=\zeta_T=1/2$~\cite{blat94}.
$t_z$ is the time needed for the line element in the $z$-th 
plane to feel the elastic interaction with the line element
in the plane $z=0$. 
In Fig.~\ref{f:wyb} we show the times scales $t_1$ for neighboring
planes separation and $t_L$ for full system size
separation. We found that for $\tau < t_1$,
$W(z,\tau) \sim \tau$ for all $z$. 
For times $\tau > t_L$, $W(L,\tau)$ saturates. 
For times $t_1<\tau<t_z$ 
the dynamic wandering shows an intermediate regime between
these two extreme behaviors.  
In Fig.~\ref{f:wyb}(b) the MSD is shown and the same regimes are
identified. First,  for $\tau < t_1$
the MSD follows the 2D diffusion 
of individual line elements (``pancakes''), $B(\tau)\sim
\tau$, before the elastic interplane interaction
becomes relevant. Second, for $t_1 < \tau < t_L$,
flux line thermal relaxation is observed, 
characterized by sublinear diffusion $B(\tau)\sim \tau^{\alpha}$,
with $\alpha=1/2$ \cite{ryu,blat94}. 
Third, for $\tau > t_L$, we observe
diffusion of the center of mass of the flux line 
$B(\tau)\sim \tau$. $t_L$ is the time
scale above which finite size effects dominate.
The time regime we wish to study is 
the one
without finite size effects, $\tau < t_L$
($\tau \lesssim 10^4$ for $L=50$).

\begin{figure}[!tbp]
\includegraphics[angle=-90,width=\linewidth,clip=true]{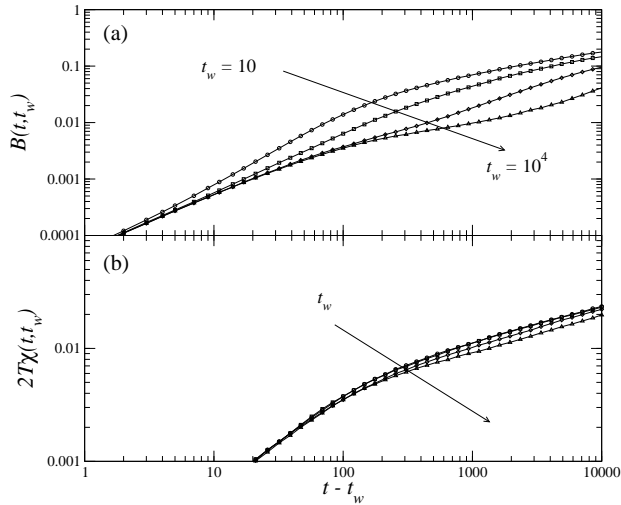}
\caption{\label{f:byx} (a) Mean square displacement and (b)
integrated response for the VG at $T=0.02$.
Different waiting times $t_w=10,10^2,10^3$ and
$10^4$ are shown, from top to bottom.}
\end{figure}

Secondly, we analyze the evolution of interacting  lines
in the presence of disorder. When we quench the system
to relatively high temperatures ($T>0.2$) it reaches equilibrium,
$W(t,t_w)$ and $B(t,t_w)$ are independent of $t_w$ and they behave
as in Fig.~\ref{f:wyb} for 
non-interacting flux lines meaning that we are clearly 
within the VL. 
Below a crossover temperature $T_g \approx 0.18$ 
the system is no longer able to equilibrate,
and the correlation functions depend on $t_w$. 
The MSD at $T=0.02$  is shown in Fig.~\ref{f:byx}(a).
For $\tau < t_1$, the dynamics is governed by single pancake  
fluctuations and the MSD is independent of $t_w$.
For $\tau > t_1$, there is ``aging'', the longer the
waiting time $t_w$, the slower the relaxation of the 
flux lines \cite{nicodemi}.

In order to study the modifications of the 
fluctuation dissipation
theorem (FDT) in the out-of-equilibrium regime,
a response function should be measured. To this end, a
random force of the form ${\bf f}_{iz}= \delta s_{iz} \hat{{\bf
x}}$ is switched on at a time $t_w$ on a replica
of the system, where $\delta$ is the intensity
of the perturbation, and $s_{iz}=\pm1$ with equal
probability \cite{parisi,kolton}. 
The integrated response is
$$\chi(t,t_w)= \frac{1}{LN\delta}
\sum_{lz} \langle s_{iz}[x^{\delta}_{iz}(t)-x_{iz}(t)]\rangle,$$
where $x^{\delta}_{iz}$ and $x_{iz}$ correspond to the
position evaluated in two replicas of the system, with and
without the perturbation. In equilibrium, FDT
implies $2T \chi(\tau) = B(\tau)$.
In Fig.~\ref{f:byx}(b), we show $2T\chi(t,t_w)$ at
$T=0.02$. $\chi$ depends on $t_w$ showing aging.
It is also clear that for long $\tau$ and long $t_w$
$2T\chi$ is not proportional to $B$, violating FDT.
This type of behavior: aging in $B$ and $\chi$ and violation
of FDT for long $t_w$ and $\tau$ is observed at all $T<T_g$.

Typically, a correlation function $C(t,t_w)$
in structural glasses has an additive scaling form
$C(t,t_w)=C_{eq}(t-t_w)+ C_{ag}[h(t)/h(t_w)]$, with, very often, 
$h(t)=t$ (known as simple aging) \cite{lesuch,crisanti}. 
This does not hold in the vortex problem. Instead,
we found a ``multiplicative'' scaling 
similar to the one proposed for the low
temperature behavior of a directed polymer in random media 
\cite{yosh98} and the out-of-equilibrium critical dynamics of the
$2d$ XY model \cite{berthier}. 
Following Yoshino~\cite{yosh98}
we tried the scaling form 
$B(t,t_w)=\tilde B(\tilde t)t_w^{\alpha}$ and 
$2T\chi(t,t_w)=\tilde \chi(\tilde t)t_w^{\alpha}$, with 
$\tilde t=t/t_w$ and $\tilde B$ and 
$\tilde \chi$ given by 
\begin{eqnarray}
\label{tildeB}
\tilde B(\tilde t)
&=&
\left\{
\begin{array}{ll} 
c_1(T)(\tilde t-1)^{\alpha(T)}
&
\qquad
\tilde t\ll1
\; , 
\\
c_2(T)(\tilde t-1)^{\alpha(T)}
&
\qquad
\tilde t\gg1
\; ,\nonumber \\ 
\end{array}
\right.
\\ 
\label{tildechi}
\tilde \chi(\tilde t)
&=&
\left\{\begin{array}{ll} 
c_1(T)(\tilde t-1)^{\alpha(T)}&
\tilde t\ll1 
\; , 
\nonumber \\ 
y(T)c_2(T)(\tilde t-1)^{\alpha(T)}&\tilde t\gg1
\; ,
\\ 
\end{array}\right.
\end{eqnarray}
where $c_1$, $c_2$ and $y$ are temperature dependent coefficients 
\cite{note}.
$y(T)$ measures the modification of FDT \cite{crisanti}
$2T\chi(t,t_w)=y(T)B(t,t_w)$ (or $\tilde \chi=y(T)\tilde B$), 
and an effective temperature \cite{peliti} 
is defined by $T_{\rm eff}=T/y$.

\begin{figure}[!tbp]
\includegraphics[width=\linewidth,clip=true]{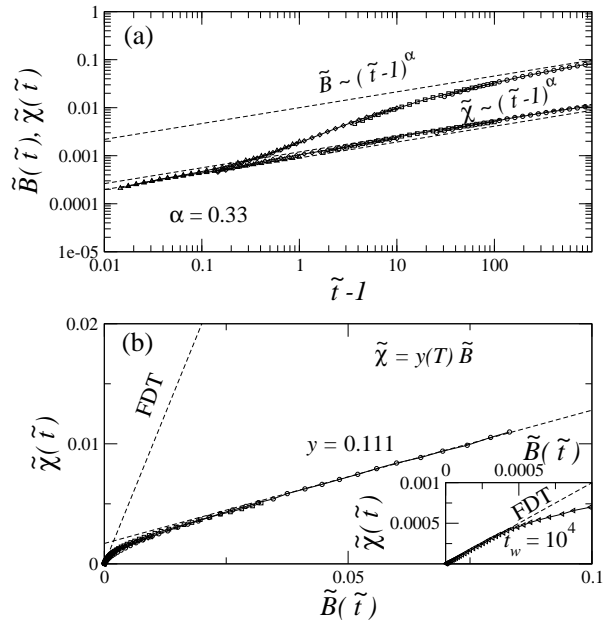}
\caption{\label{f:scypar} (a) Multiplicative scaling for the mean square
displacement and the integrated response function corresponding to 
the data in Fig.~\ref{f:byx}. $T=0.02$, and the value of $\alpha$ 
is quoted. (b) The
parametric plot $\tilde \chi(\tilde B)$ for the same data. FDT holds at
short rescaled times while a violation of FDT with $y=0.111$ 
holds at longer times. The inset shows the short rescaled time FDT
regime for $t_w=10^4$.}
\end{figure}

\begin{figure}[!tbp]
\includegraphics[width=\linewidth,clip=true]{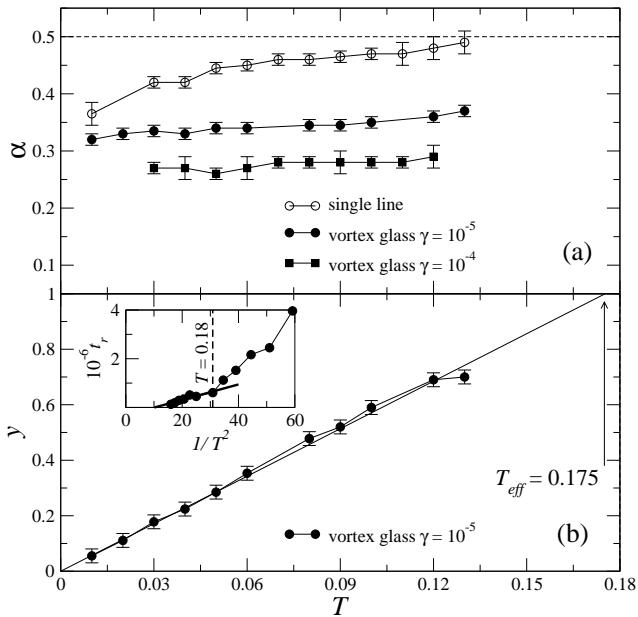}
\caption{\label{f:ayy} (a) Temperature dependence of the dynamic exponent.
The equilibrium value for single elastic lines, $\alpha =
1/2$, is highlighted. 
(b) The slope $y(T)$.
The data is fitted linearly with $y=T/T_{\rm eff}$  and the
values of $T_{\rm eff}$ are indicated.
Inset shows the relaxation time at the VL-VG
transition.
}
\end{figure}

In Fig.~\ref{f:scypar}(a) we show the scaled $\tilde B$
and $\tilde \chi$ at $T=0.02$.
The data for different $t_w$ fall on two master curves.
$\tilde B$ and $\tilde \chi$
coincide for $\tilde t \ll 1$, 
which means that FDT holds, while
for longer times ($\tilde t\gg1$) $y(T)<1$ signals the
violation of FDT. In Fig.~\ref{f:scypar}(b) a parametric plot 
of the scaled $\tilde B$
and $\tilde \chi$  is shown, for $\tilde t \ll 1$ it
has a slope equal to one, corresponding to the FDT regime (see an
enlargment in the inset),
while for longer times the slope is $y=0.111$. 
From this kind of plot the temperature dependent dynamic exponent
$\alpha(T)$ and the parameter $y(T)$ in Fig.~\ref{f:ayy},
are obtained.
At high temperatures $\alpha = 1/2$ as for single flux lines. 
The low temperature
phase has a lower dynamic exponent ($\alpha(T)<0.4$),
implying a much slower relaxation. In Fig.~4(a) we
see that $\alpha(T)$ depends weakly with $T$ and that it
decreases with increasing disorder strength
within the glassy regime. In Fig.~\ref{f:ayy}(b)
we show $y(T)=T/T_{\rm eff}$; it is well described by a linear 
form implying an effective temperature $T_{\rm eff}$
that is independent of $T$.
A linear fit yields $T_{\rm eff}=0.175$. It is remarkable
that this value is very near the crossover
temperature $T_g\approx 0.18$ below which aging is observed. 
A similar result is observed
in structural glasses: $T_{\rm eff}\approx T_g$
(as in a random energy model scenario \cite{parisi}).

To further check a ``fragile glass transition scenario'' \cite{angell},
we estimated a characteristic relaxation time
by evaluating $C(t,t_w)=\exp[-B(t,t_w)]$.
In the VL we expect 
$C(\tau) \sim \exp[-(\tau/t_r)^{\alpha}]$.
For single flux lines the $T$-dependence of $t_r$ should be
$t_r \propto 1/T^2$ \cite{ryu,blat94}. We analyzed this correlation
function for different $T$, fitting the corresponding
stretched exponential with $\alpha=1/2$ and 
estimating the relaxation time $t_r(T)$, that 
is plotted in the inset of Fig.~\ref{f:ayy}(a). 
In the VL we obtained $t_r \propto 1/T^2$, as expected.
Below  $T_g \approx 0.18$ the system does not equilibrate
but we can extract a ``$t_r$'' by fitting $C(t,t_w)$ 
for the largest $t_w$ with the same exponential form. 
The obtained $t_r$ increases very rapidly below $T_g$, 
reflecting that the system falls out of equilibrium
in this aging regime.

Finally, we performed the same analysis for
single flux lines 
in the presence of disorder. For high temperatures we found
no aging and $\alpha\approx1/2$, with important finite size effects 
[cfr.~Fig.~\ref{f:wyb}]. Below a crossover temperature
we observe similar aging to the one shown 
in Fig.~\ref{f:byx}. The multiplicative scaling,
and the same definitions of $\tilde B$ and $\tilde \chi$,
describe the data very accurately, the only 
differences being that $\alpha(T)$ takes larger values that are 
closer to $\alpha=1/2$ than in the VG [see Fig.~\ref{f:ayy}(a)], 
and that $T_{\rm eff}$ takes a 
smaller value, $T_{\rm eff}= 0.156$, than in the VG.

The results described in the last paragraph imply
that, rather surprisingly, the out-of-equilibrium
slow dynamics of the VG is dominated by the  relaxation
of the elastic lines along the $z$-direction.
Aging follows a multiplicative
scaling as for single lines, but with a smaller exponent
$\alpha(T)$.
Do vortices freeze like a structural (fragile) glass?
Yes and No. Yes, because we found aging  and
a simple violation of FDT
with an effective temperature $T_{\rm eff}$ 
which is independent of $T$ within our numerical accuracy. 
No, because aging  follows a multiplicative scaling,
similar to polymers in random media and, more precisely, critical systems
like the $2d$ XY model \cite{berthier}.
Longer times are needed to decide whether this ``critical'' scaling
holds in the limit $t_w\rightarrow\infty$ and to grasp the nature of
the low-$T$ phase.  Our results suggest to do experiments with a fast
quench to low $T$: with relaxational measurements
of transport or magnetic properties one could study the aging regime
while with voltage or flux noise measurements one could test how the
FDT is violated.

We thank discussions with H. Yoshino
and 
T. Grigera and support from SECYT-ECOS,
Conicet, CNEA, ANPCYT and ICTP grant NET-61, and 
Fundaci\'on Antorchas (SB).
LFC is a member of IUF.


\begin{thebibliography}{}
\bibitem{note} We assumed a linear dependence of $\tilde \chi$ with
 $\tilde B$, but a non trivial function $\tilde \chi(\tilde B)$,
as found in 
the spin-wave approximation to 
the  $2d$ XY model \cite{berthier}, cannot be discarded.
\end{thebibliography}
\end{document}